\documentclass[aps,prd,preprint,floatfix,superscriptaddress,nofootinbib,longbibliography,a4paper]{revtex4-1}
\pdfoutput=1

\usepackage{graphicx}
\usepackage{dcolumn}
\usepackage{bm}
\usepackage{color}
\usepackage{graphicx}
\usepackage{textcomp}
\usepackage{subfigure}
\usepackage{feynmp}
\usepackage{amsmath,amssymb,array}
\usepackage{enumerate}
\usepackage{slashed}
\usepackage{hhline}
\usepackage{hyperref}
\hypersetup{colorlinks=true,
	linkcolor=blue,
	filecolor=magenta,      
	urlcolor=blue}
\usepackage{tcolorbox}
\definecolor{bred}{rgb}{1.0,0.0,0.0}
\definecolor{bgreen}{rgb}{0.0,1.0,0.0}
\newtcbox{\rb}{on line,
  colframe=bred,colback=bred!15!white,
  boxrule=0.0pt,arc=3pt,boxsep=0pt,left=2pt,right=2pt,top=2pt,bottom=2pt}
\newtcbox{\gb}{on line,
  colframe=bgreen,colback=bgreen!15!white,
  boxrule=0.0pt,arc=3pt,boxsep=0pt,left=2pt,right=2pt,top=2pt,bottom=2pt}

\newcommand{\beqa}{\begin{eqnarray}}
\newcommand{\eeqa}{\end{eqnarray}}
\newcommand{\be}{\begin{equation}}
\newcommand{\ee}{\end{equation}}
\newcommand{\ba}{\begin{array}} 
\newcommand{\ea}{\end{array}}

\begin{document}


\title{Radiative Mass Mechanism: Addressing the Flavour Hierarchy and Strong CP Puzzle}

\author{Gurucharan Mohanta}
\email{gurucharan.phy@gmail.com}
 \affiliation{Theoretical Physics Division, Physical Research Laboratory, Navarangpura, Ahmedabad-380009, India}
\affiliation{Indian Institute of Technology Gandhinagar, Palaj-382055, India}


\begin{abstract}
We propose a class of models based on the parity invariant Left-Right Symmetric Model (LRSM), which incorporates the mechanism of radiative generation of fermion masses while simultaneously possessing the solution to the Strong CP problem. A flavour non-universal gauged abelian symmetry is imposed on top of LRSM, which helps in inducing the masses of second and first-generation fermions at one-loop and two-loop, respectively, and thereby reproduces the hierarchical spectrum of the masses. Parity invariance requires the vanishing of the strong CP parameter at the zeroth order, and the non-zero contribution arises at the two-loop level, which is in agreement with the experimental constraints. The minimal model predicts flavour symmetry breaking scale and the $SU(2)_R$ symmetry breaking scale at the same level. Flavour non-universality of the new gauge interaction leads to various flavour-changing transitions both in quarks and leptonic sectors and, therefore, has various phenomenologically interesting signatures. The model predicts a new physics scale near $10^8$ GeV or above for phenomenological consistent solutions. This, in turn, restricts strong CP phase $\bar{\theta} \lesssim 10^{-14}$ as the parity breaking scale and flavour scale are related in the minimal framework.

\end{abstract}

\maketitle
\newpage


\section{Introduction}
Fermion mass hierarchies and the strong CP, other than the Higgs mass hierarchy problem, remain the most puzzling feature of the flavour sector of the Standard Model (SM) of particle physics. Several orders of mass hierarchies of the SM fermions and their incalculability ask for a better understanding of the mass-generating mechanism. Often, this requires new symmetries and different mechanisms, making the SM a low-energy effective theory. One such mechanism that explains the origin of the hierarchical pattern is radiative mass generation. In this mechanism, the masses of only third-family fermions are non-vanishing at the zeroth order, and the lighter family masses are induced radiatively using higher-order corrections in the perturbation theory. This mechanism not only explains the origin of mass hierarchy but also makes masses partially computable parameters of the theory \cite{Weinberg:1972ws,Georgi:1972hy,Barr:1978rv,Wilczek:1978xi,Yanagida:1979gs}. The successful implementation of the mechanism requires new symmetries and/or new fields \cite{Balakrishna:1987qd,Balakrishna:1988ks,Balakrishna:1988xg,Balakrishna:1988bn,Babu:1988fn,Babu:1989tv}. The new symmetries that incorporate this mechanism are needed to be flavour-dependent (see \cite{Dobrescu:2008sz,Graham:2009gr,Endo:2019bcj,Weinberg:2020zba,Baker:2020vkh,Jana:2021tlx,Chiang:2021pma,Yin:2021yqy,Chiang:2022axu,Baker:2021yli,Mohanta:2022seo,Mohanta:2023soi,Kuchimanchi:2024nkt,Mohanta:2024wcr,Chang:2024snt,Jana:2024icm} for recent works in this direction). 

The other issue in the flavour sector is the strong CP. In a general quantum field theory, the strong CP parameter is defined as: 
\be \label{theta-bar}\bar{\theta } =\, \theta_{QCD}\, +\, Arg. (\det(M_u M_d))\,,\ee
where $M_{u (d)}$ is the mass matrix for the up (down) quark sector and $\theta_{QCD}$ is defined by the relation;
\be {\cal L}_{\theta_{QCD}} \,=\, \frac{\theta_{QCD}\, g_s^2}{32\pi^2} \, G^a_{\mu \nu}\tilde{G^a}{}^{\mu \nu}\, .\ee
This term, which is allowed by the SM gauge symmetry,  is P and T (Parity and Time reversal symmetry) violating, and therefore CP violating. However, such a term is not physical and can be absorbed in the second term of eq. (\ref{theta-bar}) or vice-versa through the redefinition of the quark fields. The physical parameter $\bar{\theta}$  appears in the neutron electric dipole moment (EDM) expression, and non-observation of neutron EDM puts a strong constraint of $\bar{\theta} < 10^{-10}$ \cite{Abel:2020pzs}. As it is a dimensionless parameter, it is theoretically anticipated to assume ${\cal O}(1)$ value like the weak phase instead of taking a tiny value. This puzzle is known as the Strong CP problem. Popular solutions include Massless quarks solutions \cite{Georgi:1981be,Choi:1988sy}, Pecci-Quinn axion solutions \cite{Peccei:1977hh,Peccei:1977ur,Wilczek:1977pj,Weinberg:1977ma,Kim:1979if,Shifman:1979if,Dine:1981rt}, Nelson-Barr class of solutions \cite{Nelson:1983zb,Barr:1984qx}, and Parity solutions \cite{Babu:1988mw,Babu:1989rb,Barr:1991qx,Craig:2020bnv}\footnote{Recent developments in the direction of parity solutions of strong CP can be found in \cite{Kuchimanchi:2010xs,Dcruz:2022rjg,Kuchimanchi:2023imj,Babu:2023srr,Babu:2023dzz,Li:2024sln,Babu:2024glr}.}. In this article, we will discuss the Parity solutions to strong CP while simultaneously accommodating the mechanism of radiative mass generation for fermions and will explore the interconnections of these two mechanisms.

In parity solutions to strong CP class of models, a Left-Right Symmetric Model (LRSM) with parity invariance imposed is considered to forbid the $\theta_{QCD}$ at the leading order. The masses of the charged fermions in this class of models are generated through the universal seesaw mechanism. The mechanism requires an extra three generations of vector-like fermions for each kind of charged fermions, and the masses of all the SM fermions are induced at the tree level itself. By making mass matrices hermitian, parity invariance forbids the tree-level $\bar{\theta}$. Small and non-vanishing strong CP phase is generated at two-loop levels, which satisfies the neutron EDM constraints. Although these models have a natural tendency to explain the smallness of  $\bar{\theta}$, the arbitrariness of the Yukawa couplings still have hierarchies of  $ \,{ Y} \sim 10^{-3}- 1$. The present work aims to construct a model in the direction of solving strong CP by considering ${\cal O}(1)$ Yukawa couplings. 

In this work, we first show that models of radiative fermion masses can also solve the strong CP problem when implemented in a parity invariant L-R symmetric theory. Here, parity invariance is used to obtain a small, strong CP phase, and the lighter generation SM fermions' masses are induced through quantum corrections. As the first two generation fermions are massless at the tree level, therefore the $Arg.\,(\det(M_u M_d))$ is vanishing at the leading order. This stands on equal footing with massless quark solutions as, at this stage, $\bar{\theta}$ is unphysical and can be completely rotated away by redefining the massless chiral fields. We impose a flavour non-universal abelian gauge symmetry $G_F$, which realises the radiative mass generation mechanism in the model. The additional fermion sector in our model is minimal compared to the other class of models along the direction of parity solutions to strong CP. We extend the fermion sector by only an extra generation of vector-like fermions for each type of charged fermions, which gives tree-level seesaw masses to the third-generation SM fermions only. Masses of the first two generation fermions are induced at one loop and two loops through the gauge corrections and, therefore, explain the observed hierarchical spectrum of masses. The subsequent loop suppression dominantly contributes to the intergenerational hierarchy and allows us to consider the ${\cal O}(1)$ values for the Yukawa couplings. The non-universal $G_F$ that suffices the above mechanism is $U(1)_{2-3}$, which is the all-fermion version of the $L_\mu - L_\tau$ symmetry. Also, it is found that mass corrections induced only by gauge boson of $G_F$ don't violate parity symmetry, and the respective strong CP phase remains vanishing at all orders of perturbation theory. However, a non-vanishing tiny $\bar{\theta}$ is generated when scalar corrections are included. The minimal version of the model predicts flavour symmetry breaking scale and the $SU(2)_R$ symmetry breaking scale at the same level. Since both are related, the flavour symmetry-breaking scale constrains the parity-breaking scale or vice-versa. As the flavour non-universality of the new flavoured gauge interactions leads to large flavour violating charges, the respective breaking scale is severely constrained. The phenomenological study of various flavour-changing transitions both at quarks and leptonic sectors, with most of them occurring at the zeroth order, has already been done in \cite{Mohanta:2022seo,Mohanta:2024wcr}.

The rest of the article is arranged as follows: The explicit model, which implements the radiative mass generation mechanism and possesses a solution to strong CP, is discussed in section \ref{sec:model}. In section \ref{sec:masses-theta}, we study the induction of fermion masses and strong CP phase at different orders of perturbation theory.  The phenomenological aspects of the model are outlined in section \ref{sec:pheno}. Then, in section \ref{sec:qualitative}, we discuss the qualitative features of the model. The study is summarised in section \ref{sec:summary}.

\section{The model}
\label{sec:model}
The gauge symmetry for the model is $SU(3)_C \times SU(2)_L \times SU(2)_R \times U(1)_{B-L} \times G_F$ with parity invariance imposed. Here $G_F= U(1)_{2-3}$ symmetry, is the generalised version of well-known $L_\mu - L_\tau$ symmetry to include all fermions. The particle contents of the model are listed in Table \ref{tab:particle} with $i=1,2,3$ representing the three generations of SM fermions. Under the new flavour non-universal symmetry $G_F$, the second and third family fermions are charged, leaving the first family neutral.  In addition to the SM fermions, a pair of vector-like fermions are added to each sector of charged fermions, which are neutral under the flavour symmetry. Three copies of the scalars $H_{Li}$ ($H_{Ri}$) is considered which transform as doublet under $SU(2)_L$ ($SU(2)_R$).
\begin{table}[!h]
\begin{center}
\begin{tabular}{ccc}
\hline
\hline
~~Particles~~ &~~ ${\cal G}_{LRSM}$~~&~~ $G_F$\\
\hline
$Q_{Li}=\left(\ba{c}u\\d \ea \right)_{Li}$ & $ \left(3,2,1,\frac{1}{3} \right)$ & ~~ $ \{0,1,-1 \}$ \\
$Q_{Ri}=\left(\ba{c}u\\d \ea \right)_{Ri}$ & $\left(3,1,2,\frac{1}{3} \right) $& ~~ $ \{0,1,-1 \}$ \\
$L_{Li}=\left(\ba{c}\nu\\e \ea \right)_{Li}$ & $ \left(1,2,1,-{1} \right)$& ~~ $ \{0,1,-1 \}$ \\
$L_{Ri}=\left(\ba{c}\nu\\e \ea \right)_{Ri}$ & $\left(1,1,2,-{1}\right)$ & ~~ $ \{0,1,-1 \}$ \\
\hline 
$H_{Li}$ & $\left(1,2,1,1 \right)$ & ~~ $ \{0,1,-1 \}$ \\
$H_{Ri}$ & $\left(1,1,2,1 \right)$& ~~ $ \{0,1,-1 \}$ \\
\hline
$U_{L,R}$ & $\left(3,1,1,\frac{4}{3} \right)$ & 0\\
$D_{L,R}$  & $\left(3,1,1,-\frac{2}{3} \right)$ & 0\\
$E_{L,R}$ & $\left(1,1,1,-2 \right)$ & 0 \\
\hline
\end{tabular}
\end{center}
\caption{Particle contents of the model.}
\label{tab:particle}
\end{table}

 The transformation properties of the above fields under the parity are defined below;
 \begin{align}
 \label{parity}
Q_{Li} & \longleftrightarrow\,  Q_{Ri} ,&\, L_{Li}\longleftrightarrow  \,  L_{Ri}\,, \nonumber\\
F_{L}&  \longleftrightarrow  \,  F_{R} ,&\, H_{Li} \longleftrightarrow  \,  H_{Ri} \,.
\end{align}
with $F= U, D, E$ representing three types of vector-like fermions with electromagnetic charges $\frac{2}{3}, -\frac{1}{3}, -1$ respectively .\\

The gauge covariant derivative can be written as: 
\be D_\mu = \begin{cases}
    \ba{c} \partial_\mu \,+\, i g W^i_{\mu L} \frac{\sigma^i}{2} \,+\, i g_1 B_\mu \frac{Y_1}{2}\, +\, i g_X X_\mu \frac{X}{2} ~~~~~~\text{For LH fields}\\ \partial_\mu \,+\, i g W^i_{\mu R} \frac{\sigma^i}{2} \,+\, i g_1 B_\mu \frac{Y_1}{2}\, +\, i g_X X_\mu \frac{X}{2}     
 ~~~~~~\text{For RH fields}\ea  
\end{cases}\,.\ee
Here $g$ , $g_1$  and $g_X$ are the  couplings for the weak $SU(2)_{L,R}$ gauge bosons,  $U(1)_{B-L}$ gauge boson and $U(1)_F$ gauge boson respectively. The coupling $g$ is made the same for both $SU(2)_L$ and $SU(2)_R$ for invariance under the parity transformation $W_{\mu L} \leftrightarrow W_{\mu R} $.  $Y_1$ and $X$ are the $B-L$ and $G_F$ quantum numbers for the fields on which the covariant derivative acts. When the scalars take vacuum expectation values (vevs), the full gauge symmetry is broken spontaneously, leaving $SU(3)_C \times U(1)_{EM}$ intact. Diagrammatically, the breaking pattern is;
\beqa SU(3)_C \times SU(2)_L \times SU(2)_R \times U(1)_{B-L} \times G_F & \xrightarrow[]{\langle H_{Ri}\rangle} &\, {SU(3)_C \times SU(2)_L \times U(1)_Y} \,\nonumber \\  & \xrightarrow[]{\langle H_{Li}\rangle}& \,  SU(3)_C \times U(1)_{EM}\, . \eeqa
The hyper-charge $Y$ and the electromagnetic charge $Q$, in this convention, are identified as:
\be \frac{Y}{2} = T^3_R \,+\, \frac{Y_1}{2}\, , ~~~~~~~ Q\,=\,T^3_L \,+\, \frac{Y}{2}\, . \ee
Denoting the vevs of the scalars as: 
\be \langle H_{Li} \rangle = v_{Li} \, ~~~ \text{and}~~~ \langle H_{Ri} \rangle = v_{Ri}\, , \label{vevs}\ee
the masses of the charged gauged bosons can be written as; 
\be M^2_{W^{\pm}_L} \,=\, \frac{1}{4} g^2 \sum_i v^2_{Li}, \,~~~\text{and}~~M^2_{W^{\pm}_R} \,=\, \frac{1}{4} g^2 \sum_i v^2_{Ri}.\ee

At the zeroth order, the charged gauged bosons don't mix with each other. However, the neutral gauge bosons mix, and the mass squared matrix for neutral gauge bosons $(W^3_L{}_\mu, W^3_R{}_\mu, B_\mu, X_\mu)$ can be parameterised as: 
\be \label{mass:gauge} {\rm M}^2 \,=\,\frac{1}{4} 
\left(\ba{cccc} g^2 \sum_i v^2_{Li} & 0 & -g\,g_1\,   \sum_i v^2_{Li} & g\,g_X (v^2_{L3}-v^2_{L2}) \\
0& g^2 \sum_i v^2_{Ri}  & -g\,g_1\,   \sum_i v^2_{Ri} & g\,g_X (v^2_{R3}-v^2_{R2}) \\
-g\,g_1 \sum_i v^2_{Li} & -g\,g_1 \sum_i v^2_{Ri} & g^2_1\,   \sum_i (v^2_{Li}\,+\,v^2_{Ri}) & g_1\,g_X \sum_P(v^2_{P2}-v^2_{P3}) \\
 g\,g_X (v^2_{L3}-v^2_{L2}) & g\,g_X (v^2_{R3}-v^2_{R2}) & g_1\,g_X \sum_{P}(v^2_{P2}-v^2_{P3})  & g^2_X (v^2_{R2}+v^2_{R3}) \ea \right)\,.
\ee
Here, the $\sum_P$ includes terms with $P=L,R$. It can be seen that the $4 \times 4$ mass matrix, ${\rm M}^2$ has a vanishing determinant (the upper-left $3\times3$ block also has determinant zero). The corresponding massless eigenstate can be identified as photon $A_\mu$  with coupling constant defined by:
\be \frac{1}{e^2}\,=\,\frac{2}{g^2}\,+\, \frac{1}{g_1^2}\, .\ee

Also, the submatrix:
\be \left( \ba{cc} g^2 \sum_i v^2_{Ri}  & -g\,g_1\,   \sum_i v^2_{Ri} \\-g\,g_1 \sum_i v^2_{Ri} & g^2_1\,   \sum_i (v^2_{Li}\,+\,v^2_{Ri}) \ea\right)\, ,\ee
has a vanishing eigenvalue in the limit $v_{Li} \to 0$. For small and non-zero $v_{Li}$, the lighter gauge boson with mass proportional to $v_{Li}$ can be identified as the SM $Z$ boson, and the heavy state with mass proportional to $v_{Ri}$ will be the $Z_{R}$ boson. From eq. (\ref{mass:gauge}), the mixing of $X$ with $Z_R$ is suppressed by a factor ${\cal O}(\frac{v^2_{R3}-v^2_{R2}}{v^2_{R3}+v^2_{R2}})$. For multi-TeV $X$ boson, the mixing with $Z$ boson will be suppressed by ${\cal O}(\frac{m^2_{Z}}{M^2_{X}})$ and satisfies all the electroweak constraints (see, for example, \cite{Mohanta:2022seo}). In this minimal setup, $H_{Ri}$ fields when assume vevs break the L-R symmetry as well as $U(1)_F$. Having the common source of breaking, the right-handed sector and the $U(1)_F$ gauge boson have masses of similar magnitude. Therefore, for further purposes, the masses of $X, Z_R$, and $W^\pm_R$ are taken of the order of $M_{X}$ i.e., the mass of $X$ boson.

\section{Fermion masses and  $\bar{\theta} $  at Leading Order and Beyond}
\label{sec:masses-theta}
\subsection{Tree level}
\label{subsec:1loop}
The imposition of well-defined parity symmetry forbids the ${\theta}_{QCD}$ term in the Lagrangian and therefore, 
the contribution to $\bar{\theta}$ arises from the second term of eq. (\ref{theta-bar}) i.e., from the phases of the quark mass matrices.  With the set of fields given in Table \ref{tab:particle}, the most general gauge and parity invariant renormalisable fermionic mass Lagrangian can be written as: 
  \beqa 
  -{\cal L}_y &=&  y^d_{i} \left(\bar{Q}_{Li}\,  {H}_{Li}\, D_{R} \, +\,\bar{Q}_{Ri} \, {H}_{Ri}\,  D_{L} \right) \,+\,y^e_{i} \left(\bar{L}_{Li}\,  {H}_{Li} \, E_{R} \,+\,\bar{L}_{Ri} \, {H}_{Ri}\,  E_{L} \right)\, \nonumber \\
  &+& y^u_{1} \left(\bar{Q}_{L1}\,  \tilde{H}_{L1} \, U_{R} \,+ \,\bar{Q}_{R1} \, \tilde{H}_{R1}\,  U_{L} \right)\, +\, y^u_{2} \left(\bar{Q}_{L2}\,  \tilde{H}_{L3} \, U_{R} \,+ \,\bar{Q}_{R2} \, \tilde{H}_{R3}\,  U_{L} \right)\, \nonumber \\ 
  &+& y^u_{3} \left(\bar{Q}_{L3}\,  \tilde{H}_{L2} \, U_{R} \,+ \,\bar{Q}_{R3} \, \tilde{H}_{R2}\,  U_{L} \right)\, \,+\, m_U\, \bar{U}_{L} U_{R } \, +\, m_D\, \bar{D}_{L} D_{R } \nonumber\\ &+& m_E\, \bar{E}_{L} E_{R }  \, +\, H.c\,.
  \eeqa

The vector-like fermions mass terms $m_{U,D,E}$ are real because of parity invariance. Also, the redefinition of various fields allows us to absorb some of the phases of $y^{u,d,e}$. It can be seen that $y^{u,e}_i$ and $y^d_3$ can be chosen real, leaving $y^d_{1,2}$ as only complex parameters of the theory. The latter generates the weak CP phase, which appears in the CKM matrix. It can also be noted that an accidental CP symmetry exists in the up-quark and the charged lepton sectors of the tree-level Lagrangian.\\

When the flavour symmetry, as well as the LR symmetry, is broken spontaneously to $SU(3)_C\times U(1)_{EM}$, the mass matrices for the charged fermions can be written as:   
  \be {\cal M}_u^{(0)} =  \left( \ba{cc} 0_{3\times 3} & \ba{c}y^u_1\, v_{L1}\\y^u_2\, v_{L3}\\y^u_3\, v_{L2}\ea \\\ba{ccc}y^u_1 v_{R1}& y^u_2 v_{R3} & y^u_3 v_{R2}\ea & m_U \ea\right),\,  {\cal M}_{d}^{(0)} =  \left( \ba{cc} 0_{3\times 3} & \ba{c}y^{d}_1\, v_{L1}\\y^{d}_2\, v_{L2}\\y^{d}_3\, v_{L3}\ea \\\ba{ccc}y^{d}_1{}^* v_{R1}& y^{d}_2{}^* v_{R2} & y^{d}_3 v_{R3}\ea & m_{D} \ea\right),  \label{Mf_0_1}\ee
  with $ v_{Li}, v_{Ri}$ are the vevs defined in eq. (\ref{vevs}) and are real.  The mass matrix ${\cal M}_{e}^{(0)}$ for charged leptons can be obtained by replacing all $y^d_i, y^d_i{}^* $ with real $y^e_i$s.  It can be seen that the above mass matrices are of Hermitian type\footnote{It can be seen that the diagonal elements of the mass matrices given in eq. (\ref{Mf_0_1}) are real. The phases of off-diagonal elements are equal and opposite to their respective transpose counterparts. This, in turn, implies a real determinant (see appendix \ref{app:scalar2}). For this reason, these matrices are called ``Hermitian Type". }, and two of the four eigenvalues are zero. Therefore, at the tree level, a massless quark solution applies, and $\bar{\theta}$ is vanishing.\\

The mass matrices given in eq. (\ref{Mf_0_1}) can be written in a compact form as: 
   \be {\cal M}_f^{(0)} =  \left( \ba{cc} 0 & \mu_f \\ \mu_f^\prime & m_F \ea\right)\, , \label{Mf_0}\ee
  with  $m_F$ as the is the vector-like fermion mass term for $F$ type, and
  \beqa \mu_u &=& \left( \ba{ccc}y^u_1 v_{L1}& y^u_2 v_{L3} & y^u_3 v_{L2}\ea \right)^T\,, ~~  \mu^\prime_u = \left( \ba{ccc}y^u_1 v_{R1}& y^u_2 v_{R3} & y^u_3 v_{R2}\ea \right)\,, \nonumber \\
  \mu_{d} &=& \left( \ba{ccc}y^{d}_1 v_{L1}& y^{d}_2 v_{L2} & y^{d}_3 v_{L3} \ea \right)^T\,, ~~  \mu^\prime_{d} = \left( \ba{ccc}y^{d}_1{}^* v_{R1}& y^{d}_2{}^* v_{R2} & y^{d}_3 v_{R3}\ea  \right) \,,\nonumber\\
  \mu_{e} &=& \left( \ba{ccc}y^{e}_1 v_{L1}& y^{e}_2 v_{L2} & y^{e}_3 v_{L3} \ea \right)^T\,, ~~  \mu^\prime_{e} = \left( \ba{ccc}y^{e}_1 v_{R1}& y^{e}_2 v_{R2} & y^{e}_3 v_{R3}\ea  \right)\,.\eeqa 

As mentioned earlier, the matrix ${\cal M}_f^{(0)}$ is of hermitian type as the complex part of   $\mu_{fi}$  is proportional to $ y^d_i$ and that of $\mu^\prime_{fi}$ is proportional to $ y^{d*}_i$. Also, it has two non-vanishing eigenvalues, which can be identified as third-generation SM fermions and vector-like fermion states. In the seesaw limit, the effective $3\times 3$ mass matrices for SM fermions can be written as: 
\be \label{M0} M^{(0)}_f = -\frac{1}{m_F}\mu_f \mu_f^\prime\, .\ee
It can also be noted that the Hermitian type structure is retained by $M^{(0)}_f$ as complex parts of $\mu_f$ and $\mu^\prime_f $ are conjugate of each other. Moreover, the above mass matrix gives masses only to third-generation fermions. The masses of lighter generation fermions arise at one-loop and two-loops through self-energy corrections induced by the flavour non-universal gauge boson as in \cite{Mohanta:2024wcr} and, therefore, can have a possible contribution to the second term of eq. (\ref{theta-bar}). The higher order corrections to $\bar{\theta}$ need to be computed only for the down-quark sector as the up-quark mass matrix given in eq. (\ref{M0}) is real and doesn't induce the complex parameters when gauge corrections are included. \\

The effects of quantum gravity are subject to the violation of all global symmetries. As parity falls under this category of symmetries, the gravity-induced corrections should be suppressed, maintaining the quality of the solution. The leading operators that could give possible contributions are: 
\be {\cal L}^{d=5} \,=\, \frac{{\cal O}(1)}{M_{Pl}}\, \bar{Q}_{Li}Q_{Rj} H^\dagger_{Rj}H_{Li}\,+ \, ..\ee
The above operators induce non-hermitian contributions to eq. (\ref{M0}). However, such contributions are much suppressed by the factor $m_F/M_{Pl}$. For $m_F \lesssim 10^{-6} M_{Pl}$, these terms don't even participate in inducing first-generation fermion masses; therefore, these corrections can be neglected.

\subsection{1-loop masses and \(\bar{\theta}\)}
\label{sec:modelB}
The 1-loop corrected fermion mass matrix can be parametrized as:
\be \label{M1}
{\cal M}_f^{(1)} = {\cal M}_f^{(0)} + \delta {\cal M}_f^{(0)}\,, \ee
where
$
\delta {\cal M}_f^{(0)}$ includes contributions from the 1-loop diagrams involving massive fermions and the $X$-gauge boson in the loop (see left panel of Fig. \ref{fig:loop}). It is evaluated as (see for example \cite{Mohanta:2022seo}):
\begin{figure}[t]
\centering
\subfigure{\includegraphics[width=0.48\textwidth]{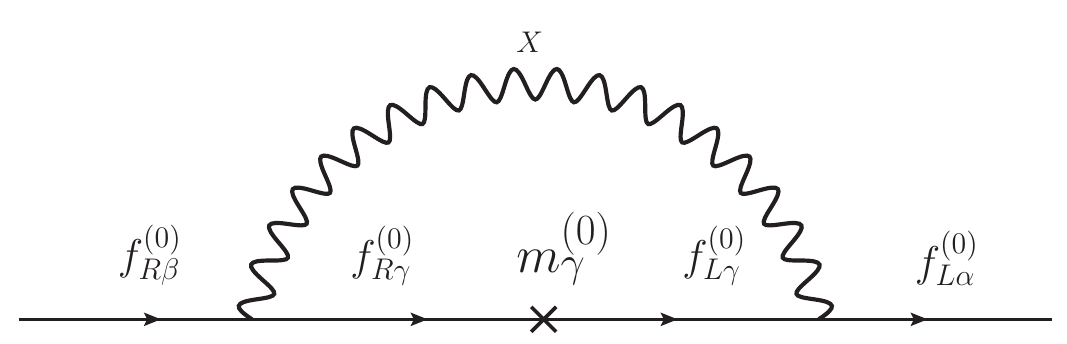}}\hspace*{0.5cm}
\subfigure{\includegraphics[width=0.48\textwidth]{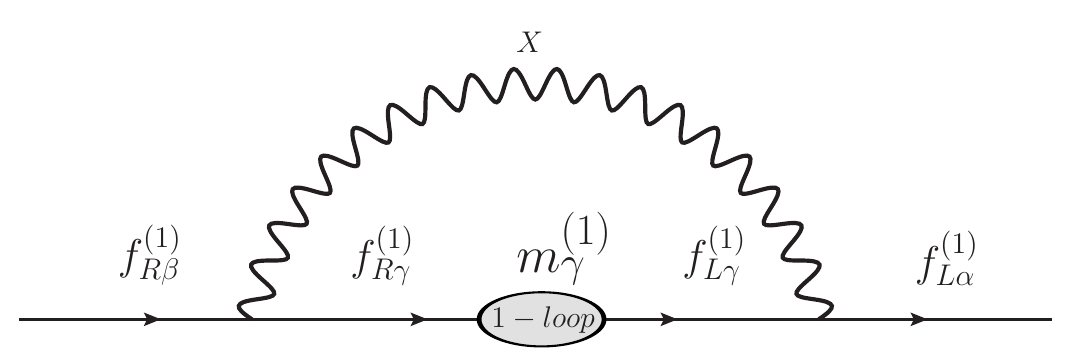}}
\caption{Loop diagrams generating the masses of second-generation fermions (left panel) and first-generation fermions (right panel).}
\label{fig:loop}
\end{figure}
\be \label{dM0_eff}
\left(\delta {\cal M}_f^{(0)}\right)_{ij} \equiv \left(\delta M_f^{(0)}\right)_{ij} \simeq \frac{g_X^2}{16 \pi^2}\, q_{Li}\, q_{Rj}\, \left(M_f^{(0)}\right)_{ij}\, \left(B_0[M_X,m_3^{(0)}] - B_0[M_X,m_4^{(0)}] \right)\,,\ee
and $\left(\delta {\cal M}_f^{(0)}\right)_{\alpha 4} = \left(\delta {\cal M}_f^{(0)}\right)_{4 \alpha} = 0$ which is because of the vector-like fermions are chosen neutral under $U(1)_F$. Here $B_0$ is the usual Passarino-Veltmann function, and $q_{Li}$ ($q_{Ri}$) is the $U(1)_F$ charges of the SM chiral field $f_{Li}$ ($f_{Ri}$). Explicitly,
\beqa
B_0[M,m] & = & \frac{2}{\epsilon}-\gamma + \ln{4\pi} + 1 -  \frac{M^2 \ln \frac{M^2}{\mu^2} - m^2 \ln \frac{m^2}{\mu^2} }{M^2 - m^2}\, ,\nonumber\\
q_{Li}&=& q_{Ri} ~\cong~ q_{i}\,=\, \lbrace 0,1,-1\rbrace\,.
\eeqa
It can be noticed that the terms proportional to $\Delta_\epsilon$ (divergent part) of $B_0$ in $\delta M^{(0)}_{ij}$ cancel, rendering the loop-induced mass finite and calculable. The resulting ${\cal M}^{(1)}$, therefore, can be written as:
\be \label{M1_1}
{\cal M}_f^{(1)} = \left( \ba{cc} \left(\delta M_f^{(0)}\right)_{3 \times 3} & \mu_{f}\\ \mu^\prime_{f} & m_F \ea \right)\,.\ee
In the seesaw limit, the unitary matrices block diagonalising ${\cal M}^{(1)} $ given in eq. (\ref{M1_1}) can be approximated as: 
\be \label{U1_ss}
{\cal U}_f^{(1)}{}_{L,R} \approx \left(\ba{cc} U_f{}_{L,R}^{(1)} & - \rho_f{}_{L,R}^{(1)} \\
 \rho_f{}_{L,R}^{(1) \dagger} U_f{}_{L,R}^{(1)} & 1 \ea\right)\,, \ee
with $\rho_f^{(1)}{}_{L,R}=\rho_f^{(0)}{}_{L,R}$ are the seesaw expansion parameters \footnote{For the present model $\rho_f^{(0)}{}_{L} = -\mu_f /m_F$  and $\rho_f^{(0) \dagger}{}_{R} = -\mu^\prime_f /m_F\,$ .  }. $U_f{}_{L,R}^{(1)}$ are the $3\times 3$ unitary matrices which diagonalises
the effective 1-loop corrected $3\times 3$ mass matrix:
\beqa \label{M1_eff}
\left(M_f^{(1)}\right)_{ij} &=& \left(M_f^{(0)}\right)_{ij} + \left(\delta M_f^{(0)}\right)_{ij}\, \nonumber \\
&=&  \left(M_f^{(0)}\right)_{ij} \left( 1+ q_{Li} q_{Rj} C \right)\, ,\eeqa
with $C=g_X^2(16\pi^2)^{-1} \left(B_0[M_X,m_3^{(0)}] - B_0[M_X,m_4^{(0)}]\right)$. It can be seen that the determinant of the above mass matrix is vanishing, implying at least one massless state. For flavour nonuniversal charges, the masses of second-generation fermions are generated, and therefore, the diagonalisation of the effective mass matrix $M^{(1)}$ can be written as: 
\be \label{M1_eff_diag}
U_f{}_L^{(1) \dagger}\, M_f^{(1)}\, U_f{}_R^{(1)} = {\rm Diag.}\left(0,m_{f2}^{(1)},m_{f3}^{(1)}\right)\,,\ee
with $m_{fi}^{(n)}$ as  mass of $i^{\rm th}$ generation fermion at $n^{\rm th}$ order. It also can be noticed from eq. (\ref{M1_eff}) that the hermitian structure of the mass matrix is retained by the one-loop corrected mass matrix and, therefore, has a vanishing contribution to $\bar{\theta}$. The vanishing of $\bar{\theta}$ can also be understood in the following way: As one of the eigenvalues of the mass matrix ${\cal M}^{(1)}$ given in eq.(\ref{M1_1}) is zero; therefore $\bar{\theta}$ is still unphysical.\\

Apart from this, X-boson induced corrections $\delta {\cal M}$, given in eq. (\ref{M1}), the mass matrix also receives corrections from the diagrams involving physical neutral scalar mixings. It can be seen that these contributions give Hermitian-type corrections to the mass matrix (see appendix \ref{app:scalar2} for detailed discussion) and don't induce the strong CP phase. However, such corrections can generate the first-generation masses, and the suppression requirement compared to second-generation fermions asks for a separate mechanism. Adding the scalar contributions also introduces several other potential parameters that are unconstrained and thereby lose the computability of the novel mechanism. However, such contributions can be made negligible by considering scalars heavy and/ or small mixing as pointed out in Ref. \cite{Mohanta:2022seo,Mohanta:2024wcr}.

\subsection{2-loop masses and $\bar{\theta}$}
\label{sec:2loop}
As mentioned earlier, the 1-loop gauge boson induced contributions generate the masses for second-generation fermions only. In this subsection, we show that 2-loop corrections can generate viable first-generation fermion masses. Following \cite{Mohanta:2024wcr}, the two-loop corrected mass matrix due to the $X$ boson induced effects (see, right panel of Fig. \ref{fig:loop}) can be parameterised in the form: 
\beqa \label{M2_eff_fnl_f}
\left(M^{(2)}_f\right)_{ij} &=& \left(M^{(0)}_f\right)_{ij} \left(1+\frac{g_X^2}{16 \pi^2}\, q_{Li}\, q_{Rj}\, (b_0[M_X,m_{f3}^{(1)}]- b_0[M_X,m_F]) \right) \nonumber \\
&+& \left(\delta M_f^{(0)}\right)_{ij} \left(1+\frac{g_X^2}{16 \pi^2}\, q_{Li}\, q_{Rj}\, b_0[M_X,m_{f3}^{(1)}] \right) \nonumber \\
&+& \frac{g_X^2}{16 \pi^2}\, q_{Li}\, q_{Rj} \,\left(U_{fL}^{(1)}\right)_{i2} \left(U_{fR}^{(1)}\right)^*_{j2}\, m_{f2}^{(1)}\, (b_0[M_X,m_{f2}^{(1)}] - b_0[M_X,m_{f3}^{(1)}])\,,\eeqa
where $U^{(1)}_{L,R}$ are the unitary matrices which are defined in eq. (\ref{M1_eff_diag}) and $b_0$ is the finite part of $B_0$ in $\overline{\rm MS}$ scheme; Thus the full mass matrix is finite. It can be seen that the first two terms are proportional to $ M_f^{(0)}$ and $\delta M_f^{(0)}$, respectively, and therefore can not generate non-hermitian entries in the mass matrix. For up-quark sector and charged lepton sector $ M_{u,e}^{(0)}$ and $\delta M_{u,e}^{(0)}$ as well as $U^{(1)}_{L,R}$ are real matrices, so  $M_{u}^{(2)}$ doesn't give contribution to $\bar{\theta}$ at two loop level. This is expected as there is accidental CP symmetry, mentioned in subsection \ref{subsec:1loop}, in the up sector at the leading order, and the gauge interactions of $X$ boson with up quarks don't violate that symmetry.\\

For the down-quark sector, although the third term proportional to $\left(U_{dL}^{(1)}\right)_{i2} \,\left(U_{dR}^{(1)}\right)^*_{j2}$ seems to induce complex entries in the mass matrix, it can be shown that they will be of hermitian nature. Here we have used the fact that $m^{(1)}_{d2}$ is real as the mass matrix $M_d^{(1)}$ is of hermitian type. $U_{dL}^{(1)}$ and $U_{dR}^{(1)}$ diagonalises the matrices $M_d^{(1)}M_d^{(1)}{}^{\dagger}$ and $M_d^{(1)}{}^{\dagger}M_d^{(1)}$, respectively.  Due to the hermitian type nature of $M_d^{(1)}$, it can be seen that the respective elements of $M_d^{(1)}M_d^{(1)}{}^{\dagger}$ and $M_d^{(1)}{}^{\dagger}M_d^{(1)}$ has same phase. This can be understood as follows. Defining $\theta_i$ as the phase of $y^d_i$, the phases of $\mu_{di}$ and $\mu^\prime_{di}$ will be $\theta_i$ and $-\theta_i$, respectively with $\theta_3=0$. Now the elements $\left(M_d^{(1)}M_d^{(1)}{}^{\dagger}\right)_{ij}$ and $\left(M_d^{(1)}{}^{\dagger}M_d^{(1)}\right)_{ij}$ can be written as:
\beqa 
\label{m2-m2}
\left(M_d^{(1)}M_d^{(1)}{}^{\dagger}\right)_{ij} &=& \frac{|\mu_{di}|\,|\mu_{dj}|}{m_F^2}\,e^{i \theta_{ij}}\,\sum_k |\mu^\prime_{dk}|^2\,\left(1+C q_{i}q_{k}\right)\,\left(1+C q_{j}q_{k}\right)\, ,\nonumber \\
\left(M_d^{(1)}{}^{\dagger}M_d^{(1)}\right)_{ij}&=& \frac{|\mu^\prime_{di}|\,|\mu^\prime_{dj}|}{m_F^2}\,e^{i \theta_{ij}}\,\sum_k |\mu_{dk}|^2\,\left(1+C q_{i}q_{k}\right)\,\left(1+C q_{j}q_{k}\right)\, .
\eeqa
Here we have used $q_{Li} = q_{Ri} \cong q_{i}$ and 
\be \label{theta_ij}\theta_{ij}=\theta_i-\theta_j\,.\ee 
The constant factor $C$ in eq. (\ref{m2-m2}) is defined in eq. (\ref{M1_eff}). From eq. (\ref{m2-m2}), it is obvious that both $\left(M_d^{(1)}M_d^{(1)}{}^{\dagger}\right)_{ij}$ and $\left(M_d^{(1)}{}^{\dagger}M_d^{(1)}\right)_{ij}$ has the same phase, and their elements can be obtained from each other by the interchange $|\mu_i| \leftrightarrow |\mu^\prime_i|$. As the phases of the earlier two are the same, therefore the phases of elements $\left(U_{dL}^{(1)}\right)_{ij}$ and $\left(U_{dR}^{(1)}\right)_{ij}$ will be the same (see, appendix \ref{app:phases} for details). It can be seen that the product $\left(U_{dL}^{(1)}\right)_{i2} \left(U_{dR}^{(1)}\right)^*_{j2}$ in eq. (\ref{M2_eff_fnl_f}) will have the phase factor $e^{i\theta_{ij}}$. Also, the phase factors of $(M_d^{(0)})_{ij}$ and $(\delta M_d^{(0)})_{ij}$ will be $e^{i\theta_{ij}}$ as both are proportional to $\mu_{di} \mu^\prime_{dj}$. Therefore, from eq. (\ref{M2_eff_fnl_f}) the phase factor of $\left(M^{(2)}_f\right)_{ij}$ will be $e^{i\theta_{ij}}$ and same for $\left(M^{(2)}_f\right)_{ji}$ will be $e^{-i\theta_{ij}}$. This implies the hermitian nature of the mass matrix is retained, and the determinant is real (similar to eq. (\ref{real-det})). So, 
\be \label{theta:2} \bar{\theta}_{\rm 2-loop}^{(X)}\,=\,Arg. {\det} (M^{(2)}_d) \,=\, 0\, .\ee

\subsection{Weak CP contribution to \(d^e_n\)}
Herein, we compute the neutron electric dipole moment, $d^e_n$, induced at the 1-loop level due to the weak CP violating phases. It can be seen that attaching a photon line to the internal fermion lines of the left panel of Fig. \ref{fig:loop} gives a possible contribution to $d^e_n$. This contribution is proportional to the imaginary part of $\delta M^d_{11}$. For the X-boson exchange diagram, $\delta M^d_{11}$ is real at 1-loop level (see eq. (\ref{dM0_eff})) and the same for the scalar mediated contribution given in eq. ({\ref{dm1:scalar}}). Therefore; 
\be d^e_n ~\propto ~ {\rm Im}\left((\delta M^{(X)}_d)_{11}\,+\,(\delta M^{(S)}_d)_{11}\right)\,\,= \, 0 \,. \ee

This is an interesting feature of our model, which results from the fact that vector-like fermion masses are real. This feature is not shared by the other class of models that use the universal seesaw for generating fermion masses by considering vector-like fermion mass terms as complex parameters. Although in their model, non-vanishing $d^e_n$ is induced at the 1-loop level, the value is small enough to fall under the experimentally allowed region.

\section{Phenomenological analysis}
\label{sec:pheno}

As previously noted in last paragraph of section \ref{sec:model}, the model features both flavour symmetry breaking and left-right symmetry breaking occurring at a similar scale. Unlike other new particles, the \( X \) boson exhibits significant flavour-violating interactions with SM fermions, making it highly relevant for phenomenological investigations. Assuming the vector-like fermions are heavier than the \( X \) boson, with their coupling additionally suppressed by the see-saw expansion parameters \( \mu/m_F, \mu'/m_F \); we focus on the flavour-changing neutral current (FCNC) effects mediated by the \( X \) boson for the remainder of this section.\\

The \( X \) boson’s flavour-violating couplings naturally induce meson-antimeson oscillations at tree level. In the lepton sector, exchange of the \( X \) boson prompts flavour-violating processes such as \( \mu \to e \) conversion in nuclei and trilepton decays \( l_i \to 3 l_j \) at the leading order. Processes like \( l_i \to l_j \gamma \) arise at the one-loop level. In our previous study (ref. \cite{Mohanta:2022seo}), a detailed phenomenological analysis of these flavour-violating effects was conducted, indicating a new physics scale around \( M_X \approx 10^8 \) GeV, consistent with all flavour constraints. A comparable bound is found in another study (ref. \cite{Mohanta:2024wcr}) for the case of \( \epsilon = 1 \) in the model discussed there. For this new physics scale, the contribution to electro-weak precision observables arising from $Z-X$ mixing and $Z-Z_R$ mixing are automatically satisfied \cite{Mohanta:2022seo}. Also, such constraint is much more stringent compared to the $W_R, Z_R $ mass constraint obtained in \cite{Craig:2020bnv}.

\section{Qualitative analysis}
\label{sec:qualitative}
The vanishing of new gauge correction induced $\bar{\theta}$ at 1-loop and 2-loop level can be guessed as the gauge interactions of $X$ don't violate the parity symmetry. This can be easy to identify in the flavour basis. For example,
\be \label{L_gauge:4}
-{\cal L}_{\rm gauge} = \frac{g_X}{2} X_\mu \left(q_{L i}\, \overline{f}_{L i} \gamma^\mu f_{L i} + q_{R i}\, \overline{f}_{R i} \gamma^\mu f_{R i} \right)\,, \ee
For the present model, $q_{Li} = q_{Ri} \cong q_{i}$, therefore the above interaction is invariant under the parity transformations defined in eq. (\ref{parity}).\\

In the mass basis of fermions, when all three generation fermion's masses are induced perturbatively, the gauge Lagrangian from Eq. (\ref{L_gauge:4}) can be rewritten as:  
\be \label{L:mass:4}
-{\cal L}_{\rm gauge} = \frac{g_X}{2} X_\mu \left((Q_f^{(2)}{}_{L})_{ ij}\, \overline{\textbf{f}}_{L i} \gamma^\mu {\textbf{f}}_{L j} + (Q_f^{(2)}{}_{R})_{ ij}\, \overline{\textbf{f}}_{R i} \gamma^\mu {\textbf{f}}_{R j} \right)
\ee
where  
\be
{ Q}_f^{(2)}{}_{L,R} = { U_f^{(2)}}_{L,R}^\dagger\,q\,{U_f^{(2)}}_{L,R}\,.
\ee
Here, \({U_f^{(2)}}_{L,R}\) are unitary matrices that diagonalise the two-loop corrected effective mass matrix from Eq. (\ref{M2_eff_fnl_f}).  \\

In the up-quark sector, these matrices are real and orthogonal, ensuring that \({ Q}_u^{(2)}{}_{L,R}\) does not introduce complex phases in the Lagrangian. In the down-quark sector, following the reasoning in Section \ref{sec:2loop} and Appendix \ref{app:phases}, it can be shown that both \(\left({U_d^{(2)}}_{L}\right)_{ik}\) and \(\left({U_d^{(2)}}_{R}\right)_{ik}\) share a common phase factor, \(e^{i \theta_{ik}}\).  
With this, the gauge Lagrangian for the down-quark sector given in eq. (\ref{L:mass:4}) takes the form:  
\be
-{\cal L}_{\rm gauge} = \frac{g_X}{2} X_\mu \left(\lvert(Q_d^{(2)}{}_{L})_{ ij}\rvert\,e^{i\theta_{ij}} \,\overline{\textbf{d}}_{L i} \gamma^\mu {\textbf{d}}_{L j} + \lvert(Q_d^{(2)}{}_{R})_{ ij}\rvert\,e^{i\theta_{ij}} \, \overline{\textbf{d}}_{R i} \gamma^\mu {\textbf{d}}_{R j} \right).
\ee
This equation highlights that the left-handed and right-handed couplings of \(X\)-boson carry the same phase. Since mass terms arise from mixing both chiralities and the mass insertion \(m^{(2)}_f\) is real, the overall product remains hermitian type. For instance, considering a general three-loop diagram (Fig. \ref{fig:xloops}), the complex component of the diagram corresponds to:
\begin{figure}[t]
\centering
\includegraphics[width=0.8\textwidth]{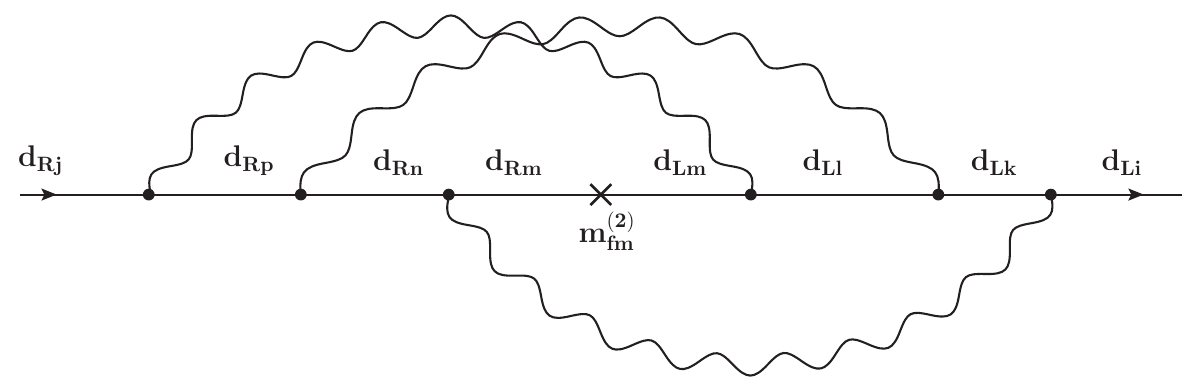}
\caption{Gauge boson induced next order corrections.}
\label{fig:xloops}
\end{figure}
\[
(Q_d^{(2)}{}_{L})_{ ik}(Q_d^{(2)}{}_{L})_{ kl}(Q_d^{(2)}{}_{L})_{ lm}(Q_d^{(2)}{}_{R})_{ mn}(Q_d^{(2)}{}_{R})_{ np}(Q_d^{(2)}{}_{R})_{ pj}
\]
which simplifies to  
\be \label{theta1}
\text{Constant} \times e^{i\theta_{ik}}\,e^{i\theta_{kl}}\,e^{i\theta_{lm}}\,e^{i\theta_{mn}}\,e^{i\theta_{np}}\,e^{i\theta_{pj}} = \text{Constant} \times e^{i\theta_{ij}}.
\ee
The above equality is obtained by using eq. (\ref{theta_ij}). Using eq. (\ref{theta1}) and Fig. \ref{fig:xloops}, we find that the phase factor of the \(ij\)-th element of the loop-corrected mass matrix is proportional to \(e^{i\theta_{ij}}\), just like \(M^{(2)}_{ij}\) in Eq. (\ref{M2_eff_fnl_f}). Thus, regardless the order of perturbation, the corrected mass matrix remains hermitian type, ensuring a real determinant as shown in Eq. (\ref{real-det}). Consequently, we conclude that gauge corrections from the \(X\) boson do not introduce a strong CP phase at any order of perturbation theory.\\

The above argument can also be understood by comparing the model structure with \cite{Babu:1989rb}. Writing the Lagrangian of our model in the form:
\be {\cal L} \,\supset\, {\cal L}_{\rm Yukawa} \,+\, {\cal L}_{X}\,+\, {\cal L}_{\rm scalar}\,,\ee
It can be seen that the part ${\cal L}_{\rm Yukawa} \,+\, {\cal L}_{X}$ doesn't violate the parity symmetry and mimics the role of ${\cal L}_{\rm Yukawa}$ with Universal seesaw mechanism of \cite{Babu:1989rb}. Therefore, the corrections induced by ${\cal L}_{X}$ don't generate $\bar{\theta}$ at all orders of perturbation theory. Parity symmetry is broken softly in the scalar sector, so the radiative effects of scalars give a potential non-vanishing strong CP phase. Following \cite{Babu:1989rb}, scalar induced 2-loop $\bar{\theta}$ can be approximated as: 
\be \bar{\theta}\,\sim\, \left(\frac{1}{16\pi^2}\right)^2 \, \left(\frac{v_L}{v_R}\right)^2\, \phi^2\, .\ee
Here, the phase $\phi$ can be originated from mixing various $\mu$ and $\mu^\prime$. The factor $v_L/v_R$ is due to the mixing of $H_{Li}$ and $H_{Rj}$. For $\phi$ as ${\cal O}(1)$,  $v_L$ as electroweak scale and $v_R$ as the $U(1)_F$ breaking scale i.e, $v_R \sim M_X \sim 10^8$, it can be seen $\bar{\theta} \sim 10^{-14}$. This value is three orders of magnitude smaller than the current experimental limit. Therefore, the model provides a viable solution to the strong CP problem.

\section{Summary}
\label{sec:summary}
The mechanism of radiative generation of fermion masses not only has a tendency to reproduce the observed spectrum of hierarchical fermion masses but also makes the latter computable parameters of the theory. The model presented in this study accommodates the said mechanism and also has a potential solution to the well-known strong CP problem. A parity invariant L-R symmetric model is employed to achieve a small, strong CP phase and the radiative mass generation for the SM fermions is incorporated by extending the gauge sector by a flavour-dependent Abelian gauge symmetry, \( G_F \). An additional generation of vector-like fermions, which are crucial for the mechanism, are added for each charged fermion type. This setup produces tree-level seesaw masses for the third-generation SM fermions, while the masses of the first two generations are generated at one- and two-loop levels via gauge corrections, allowing for \({\cal O}(1)\) values in the Yukawa couplings. The flavour non-universal symmetry \( G_F \) required for this mechanism is \( U(1)_{2-3} \), analogous to an all-fermion version of the \( L_\mu - L_\tau \) symmetry. The feature of successive increase of the rank of the mass matrices at different loops is also exploited to get a smaller strong CP phase at the higher order. It is found that the  ${\cal L}_X$ gauge corrections alone don't break parity symmetry, and the induced strong CP phase is vanishing for all orders of perturbation theory. However, scalar-induced corrections contribute to the non-vanishing $\bar{\theta}$ and at the 2-loop level, estimated to be the order of $10^{-14}$. Precise measurement of $\bar{\theta}$ may falsify the model depending upon the measured value.  \\

The minimal version of the model suggests that both the flavour symmetry breaking scale and the \( SU(2)_R \) symmetry (parity) breaking scale are comparable. The non-universality of the new flavoured gauge interactions introduces various flavour-changing transitions in both quark and lepton sectors, leading to a range of phenomenologically intriguing signatures. From various flavour-violating processes, the new physics scale is constrained to be near $10^8$ GeV or above. Compared to this constraint, the parity-breaking constraints are negligible. Although this large separation from the electroweak scale seems ugly, it is inherent in this type of class of models, which explains the mass hierarchy using gauge corrections. Moreover, such a fine-tuned solution is possible as numerous undetermined parameters exist in the scalar potential.

\begin{acknowledgments}
I sincerely thank Dr. Ketan M. Patel for his invaluable discussions, insightful suggestions, and meticulous review of the manuscript. I am also grateful to Dr. Saurabh K. Shukla for reading the draft and providing valuable feedback. This research is supported by the Department of Space (DoS), Government of India.
\end{acknowledgments}
\appendix
\section{Scalar Potential}
\label{app:scalar}

The gauge and parity invariant renormalisable  scalar potential can be written as:
\beqa \label{potential}
V &=& \mu_{L i}^2\, H_{L i}^\dagger H_{L i}\, +\, \mu_{Ri}^2\, H_{R i}^\dagger H_{R i}\, +\,   (\lambda)_{ij}\,\left[ (H_{L i}^\dagger H_{L i})\,( H_{L j}^\dagger H_{L j})\,+\, (H_{R i}^\dagger H_{R i})\,( H_{R j}^\dagger H_{R j})\,\right] \nonumber \\
& + &  (\tilde{\lambda})_{ij}\,\left[ (H_{L i}^\dagger H_{L j})\,( H_{L j}^\dagger H_{L i})\,+\, (H_{R i}^\dagger H_{R j})\,( H_{R j}^\dagger H_{R i})\,\right] \,+\,   (\lambda^4)_{ij}\, (H_{L i}^\dagger H_{L i})\,( H_{R j}^\dagger H_{R j})\nonumber \\
&+&  (\tilde{\lambda^4})_{ij} \,(H_{L i}^\dagger H_{L j})\,( H_{R j}^\dagger H_{R i}) \,,\eeqa 
with $(\,)$ bracket indicating weak singlets. Here $\mu_{L i}^2 \neq \mu_{R i}^2$ is chosen to break the parity symmetry softly. Also, $\tilde{\lambda}$ and $\tilde{\lambda^4}$ can have vanishing diagonal elements without loss of generality. It can be seen that all the potential parameters are real:
\begin{itemize}
    \item $\mu_{L,R i}^2, \lambda, \tilde{\lambda} $ and ${\lambda^4}$  are real as the corresponding operators are self-conjugate.
    \item $(\tilde{\lambda^4})_{ij}$ is real as the conjugate of its respective operator is the same as its parity-transformed one.
\end{itemize}   As all the parameters of the potential are real, we assume that they lead to real vacuum expectation values. The latter is completely deterministic from the potential parameters. Denoting vevs as $v_{L,R i}$, given in eq. (\ref{vevs}), the minimisation of the potential gives;
\be \frac{\partial V}{\partial v_{Li}} \,=\, 0 \implies 2 \mu^2_{Li}\,v_{Li}\,+\, 2 (\lambda_{ij} \,+\, \tilde{\lambda}_{ij})\,v_{Li} v_{Lj}^2\,+\, 2 \left({\lambda^4}\right)_{ij}\,v_{Li} v_{Rj}^2 \,+\,  \left(\tilde{\lambda^4}\right)_{ij}\,v_{Lj} v_{Ri}v_{Rj} = 0\,.
\ee
A similar equation holds for $L \to R$. Considering small mixing between $H_{Li}$ and $H_{Ri}$ i.e., $\lambda^4, \tilde{\lambda^4} \ll 1$, the above equations can be written as: 
\beqa  \mu^2_{Li}\,+\, (\lambda_{ij} \,+\, \tilde{\lambda}_{ij})\,v_{Lj}^2\,&=&\, 0 \,,\nonumber \\
\mu^2_{Ri}\,+\, (\lambda_{ij} \,+\, \tilde{\lambda}_{ij})\,v_{Rj}^2\,&=&\, 0\,.\eeqa
On solving,
\be \frac{v^2_{Li}}{v^2_{Ri}} \,=\,\frac{\left ((\lambda \,+\, \tilde{\lambda} )^{-1}. \mu^2_L\right )_i }{\left ((\lambda \,+\, \tilde{\lambda} )^{-1}. \mu^2_R\right )_i }\,,\ee
where $\mu^2_{L,R}=\left( \ba{ccc} \mu^2_{L1,R 1} & \mu^2_{L2,R 2}& \mu^2_{L3,R 3} \ea\right)^T$. The large separation between $v_{Li}$ and $v_{Ri}$ can be obtained only at the cost of tuning the parameters $\mu^2_{Li,Ri}$. Assuming the $v_{R}$ as the $U(1)_F$ flavour symmetry breaking scale, the degree of fine-tuning required is $\Delta^{-1} = v^2_L/v^2_R \sim 10^{-12}$. This hierarchy problem is intrinsic to the radiative mass models and is analogous to the gauge hierarchy problem.\\

Denoting $h_{Li}$ and $h_{Ri}$ as the electrically neutral components of $H_{Li}$ and $H_{Ri}$, their $6 \times 6$ mixing matrix can be expressed as: 
\be M^2_h \,=\, \left(\ba{cc} m^2_{LL} & m^2_{LR}\\m^2_{RL} & m^2_{RR} \ea \right)\,.\ee
This is a real symmetric matrix with entries  $m^2_{PP^\prime}$ ($P,P^\prime=L,R$) each of $3\times 3$ dimension. It can be diagonalised by a real orthogonal matrix ${\cal R}$, which can be written in the form;
\be {\cal R}\,=\, \left(\ba{cc} {\cal R}_{LL} & {\cal R}_{LR}\\ {\cal R}_{RL} & {\cal R}_{RR} \ea \right)\,.\ee
The physical neutral scalars $S_a$, which are obtained using the above transformation, are of the form
\be S_a \,=\, {\cal R}^T_{ab}\, h_b\,, \ee
with $h_a\,=\, ( \ba{cc} h_{Li} & h_{Ri}\ea)^T$ and $a,b=1,2,..,6$.

\section{Scalar 1-loop contributions and real determinant}
\label{app:scalar2}
The interaction of physical neutral scalars with the fermions, for example, $d$ quarks, in the mass basis will be  written as: 
\be -{\cal L}_Y \,=\, (\tilde{y}_d)_{ia}\, \bar{d^\prime}_{Li}\, S_a\, D^\prime_R\,+\,(\tilde{y}^\prime_d)_{ia}\, \bar{D^\prime}_{L}\, S_a\, d^\prime_{Ri}\, +\, {\cal O}\left(\frac{\mu_d}{m_D}, \frac{\mu^\prime_d}{m_D}\right)\,,\ee
with, 
\beqa (\tilde{y}_d)_{ia} &=& \sum_j \left(U^{(0)\dagger}_{dL}\right)_{ij}\,y^d_j\,({\cal R}_{L})_{ ja}\, , \nonumber \\
(\tilde{y}^\prime_d)_{ia} &=& \sum_j \left(U^{(0) T}_{dR}\right)_{ij}\,y^d_j{}^*\,({\cal R}_{R})_{ ja}\, .
\eeqa
In the above, ${\cal R}_{L}= (\ba{cc} {\cal R}_{LL}&{\cal R}_{LR}\ea)$ and ${\cal R}_{R}= (\ba{cc} {\cal R}_{RL}&{\cal R}_{RR}\ea)$ are the $3\times6$ submatrices of ${\cal R}$. The presence of the above interactions also contributes to the self-energy corrections of the fermions at the 1-loop level, which is diagrammatically similar to the one on the left panel of Fig. \ref{fig:loop} with $X$ boson replaced by physical scalars $S_a$. The amplitude of this diagram has the form:
\be \sigma^{(S)}_{ij}\,=\, -\frac{m_D}{16\pi^2}\, \sum_a (\tilde{y}_d)_{ia}\,(\tilde{y}^\prime_d)_{ja}\, B_0[m_{Sa},m_D]\, ,\ee
with $B_0[m_{Sa},m_D]$ as two-point Passarino-Veltman function. Corrections to the mass matrix can be written as; 
\be \label{dm1:scalar}(\delta M^{(S)})_{ij}= \left(U^{(0)}_{dL}\,\sigma^{(S)}\, U^{(0)\dagger}_{dR} \right)_{ij}\,=\, -\frac{m_D}{16\pi^2}\, {y^d_{i}y^{d*}_{j}}\sum_a  \,({\cal R}_{L})_{ ia}\,({\cal R}_{R})_{ ja} B_0[m_{Sa},m_D]\,.\ee
It can be seen that the above contribution is of hermitian nature as it is proportional to ${y^d_{i}y^{d*}_{j}}$. Also, these corrections, when added to the mass matrix given in eq. (\ref{M1_eff}) can potentially generate the first-generation fermion masses at 1-loop level only. However, such corrections can be ignored under the circumstances mentioned at the end of section \ref{sec:modelB}.\\

Including the scalar 1-loop contributions given in eq. (\ref{dm1:scalar}), the elements of the total 1-loop correct mass matrix can be written as: 
\beqa 
\label{M1-total}
\left(M^{(1)}\right)_{ij} &=& \left(M^{(0)}\right)_{ij}\,+\, \left(\delta M^{(0)(X)}\right)_{ij} \,+\, \left( \delta M^{(0)(S)}\right)_{ij}\, \nonumber\\
&=& -\frac{1}{m_F}\, \left(\mu_i \mu^\prime_j  \,+\, C \mu_i \mu^\prime_j \,q_i q_j \,+\,  \mu_i \mu^\prime_j \,A_{ij}\right)\, ,\eeqa
with the real constant factor $C$ defined in eq. (\ref{M1_eff}) and 
\be\label{Aij} 
A_{ij}\,=\,-\frac{m_D^2}{16\pi^2}\,\frac{1}{v_{Li}v_{Rj}}\sum_a  \,({\cal R}_{L})_{ ia}\,({\cal R}_{R})_{ ja} B_0[m_{Sa},m_D]\,.
\ee

The absolute hermiticity of the above matrix would require $\mu_i = \mu^\prime_i$ and $A_{ij} = A_{ji}$ (since $A_{ij}$ are real). Since, as already mentioned, the phase of $\mu_i $ and $\mu^\prime_i$ are equal and opposite, the phase of $M^{(1)}_{ij}$ and $M^{(1)}_{ji}$ turn out to be equal and opposite. Writing $M^{(1)}_{ij}$ in the form
\be M^{(1)}_{ij}\,=\, r_{ij}\, e^{i\theta_{ij}} \ee
with all $r_{ij}$ real, the determinant of $M^{(1)}$ can be written as 
\beqa
\det M^{(1)}&=& \epsilon_{ijk}\, M^{(1)}_{1i}\,M^{(1)}_{2j}\,M^{(1)}_{3k}\, \nonumber \\
&=& \epsilon_{ijk}\, r_{1i}\,r_{2j}\,r_{3k}\,e^{i(\theta_1+\theta_2+\theta_3)}\,\,e^{-i(\theta_i+\theta_j+\theta_k)}\eeqa
Here we have used the fact $\theta_{ij}=\theta_i\,-\,\theta_j$. It can be seen that the factor $(\theta_i+\theta_j+\theta_k)$ is always $(\theta_1+\theta_2+\theta_3)$ for $i\neq j\neq k$. This implies the determinant is always real, and 

\be \label{real-det} \det M^{(1)}\,=\, \epsilon_{ijk}\,r_{1i}\,r_{2j}\,r_{3k}\,.\ee

\subsection{Notes on Hermitian Type matrices} 
\begin{itemize}
    \item The simplest form of hermitian type matrix $X$ can be formed if the elements $X_{ij}$ has the phase factor $e^{i\phi_{ij}}$ with $\phi_{ij}=\phi_i -\phi_j$. The determinant is real, similar to eq. (\ref{real-det}).
    \item The sum of two hermitian type matrices $X$ and $Y$ is also a hermitian type matrix only when the phase factor of the elements $X_{ij}$ and $Y_{ij}$ are the same. 
\end{itemize}


\section{Phases of $U_{L,R}$}
\label{app:phases}
Eq. (\ref{m2-m2}) can be rewritten as: 
\beqa \label{m2-m2:2} 
\left(M M^{\dagger}\right)_{ij} &=& c_{ij}\, e^{i\theta_{ij}}\, , \nonumber\\
\left(M^{\dagger}M\right)_{ij} &=& d_{ij}\, e^{i\theta_{ij}}\,,\eeqa
where $c_{ij}$ and $d_{ij}$ are the real constants, and $\theta_{ij}$ is defined in eq. (\ref{theta_ij}). It can be seen that $c_{ij}$ and $d_{ij}$ are elements of real symmetric matrices. Similar form can be obtained for $M^{(0)}M^{(0)\dagger}$ and $M^{(0)\dagger}M^{(0)}$. In fact, the above statements and eq. (\ref{m2-m2:2}) are correct for any hermitian type matrix whose $ij$'th elements phase factor can be written as $e^{i\theta_{ij}}$. So, we have suppressed the super-scripts and subscripts for brevity. Now, inverting the bi-unitary diagonalisation equations, we get
\beqa\label{m2-m2:3} 
\left(M M^{\dagger}\right)_{ij} &=& \left(U_L \,D^2 \,U^\dagger_L\right)_{ij} \,,  \nonumber\\\left(M^{\dagger}M\right)_{ij} &=& \left(U_R\, D^2 \,U^\dagger_R\right)_{ij}  \,,\eeqa
with $D= {\rm Diag.} (m_k)$. Now, putting eq. (\ref{m2-m2:2}) in eq. (\ref{m2-m2:3}) and using the fact $\theta_{ij}=\theta_i -\theta_j$, we can write 
\beqa\label{m2-m2:4}  c_{ij} &=& \left(e^{-i\theta_{ik}}\,(U_L)_{ik}\right)\, m_k^2 \,\left(e^{i\theta_{jk}}\,(U_L)^*_{jk}\right)\,  \,, \nonumber\\
d_{ij} &=& \left(e^{-i\theta_{ik}}\,(U_R)_{ik}\right)\, m_k^2 \,\left(e^{i\theta_{jk}}\,(U_R)^*_{jk}\right) \,,\eeqa

Defining: \be\left(S_{L,R}\right)_{ik}= e^{i\theta_{jk}}\,(U_{L,R})_{jk}\,,\ee the above equation (\ref{m2-m2:4}) can be rewritten as: 
\beqa\label{m2-m2:5}  c_{ij} &=& \left(S_L\right)_{ik}\, m_k^2 \,\left(S_L\right)^*_{jk}\,  \,, \nonumber\\
d_{ij} &=& \left(S_R\right)_{ik}\, m_k^2 \,\left(S_R\right)^*_{jk}\,,\eeqa
Since  $m_k^2$ is real and, $c_{ij}$ and $d_{ij}$ are elements of a real symmetric matrix, already mentioned after eq. (\ref{m2-m2:2}), the diagonalising matrix $S_{L,R}$ has to be real orthogonal matrix. Therefore, 
\beqa
(U_L)_{ik}&=& a^L_{ik}\,e^{i\theta_{ik}}\,,\nonumber\\
(U_R)_{ik}&=& a^R_{ik}\,e^{i\theta_{ik}}
\eeqa
with $a^{L,R}_{ik}$ as any real elements. Hence it is shown that the phase of $(U_L)_{ik}$ and $(U_R)_{ik}$ are equal.

\bibliography{apssamp}

\end{document}